\documentclass[a4paper, fleqn, leqno, 12p]{article}
\usepackage{color,amsmath,amssymb,soul,graphicx}
\usepackage[footnotesize]{caption} 
\usepackage[comma,square, numbers, sort&compress]{natbib} 
\begin{document}
\title{Emergence and Maintenance of Excitability: Kinetics over Structure}
\author{Shimon Marom\thanks{Department of Physiology \& Biophysics, Faculty of Medicine, Technion--Israel Institute of Technology, Haifa 32000, ISRAEL. Email: marom@technion.ac.il. This research is funded by an Israel Science Foundation grant (1694/15).}}

\date{}

\maketitle

\begin{abstract}
\noindent Emergence and maintenance of excitability is often phrased in terms of arriving at and remaining about a manifold of `solutions' embedded in a high dimensional parameter space. Alongside studies that extend traditional focus on control-based regulation of structural parameters (channel densities), there is a budding interest in self-organization of kinetic parameters. In this picture, ionic channels are continually forced by activity in-and-out of a large pool of states not available for the mechanism of excitability. The process, acting on expressed structure, provides a bed for generation of a spectrum of excitability modes. Driven by microscopic fluctuations over a broad range of temporal scales, self-organization of kinetic parameters extends the metaphors and tools used in the study of development of excitability.   

\end{abstract}

{\small
\subsubsection*{ Highlights}
\begin{itemize}
\item Excitability entails tuning of both structure and kinetics in a large parameter space
\item Controlled protein expression places a cell in a ballpark of excitability class
\item Given expressed structure, self-organized kinetic parameters regulate excitability
\item Self-organization of kinetic parameters is driven by microscopic fluctuations
\item Controlled expression and self-organized kinetics cover a continuum of timescales
\end{itemize}
 }  
 
Over half a century of extensive scientific work teaches us that the phenomenon of excitability in biological membranes entails tuning of relations in a large parameter space. A balance of effective expression of ionic channel proteins having unique gating kinetics is required for the action potential phenomenon to emerge and be maintained. What is the nature of the processes that constrain the relations between parameters in the large combinatorial space of all possible configurations? While our understanding of the physics underlying membrane excitability is advanced compared to other physiological phenomena, we are in the dark when pushed to the corner with this question.  Allegedly, even the basic Hodgkin-Huxley (HH) formal description of excitability \cite{Hodgkin:1952jt} -- with only two types of voltage-gated ionic conducting proteins -- is not easily tamed when the kinetic parameters of channel gating are modified (which occurs due to a rich protein state space and modulatory processes); or, when the ratio between the number of different channel proteins in the membrane is changed (due to differential protein turnover); or, when the membrane capacitance and current leak change (during massive cell growth, movement or contact of the cell with biological matrices that impact on membrane surface tension). The natural phenomenon of excitability is resilient to such changes, or (at least) apparently more robust than the formal, mathematical models used to describe it.

The problem, it is acknowledged, goes beyond the regulation of excitability; it belongs to a class of open questions that concern organization in biological systems in general, the emergence of macroscopic functional `order' from a large space of potential microscopic `disordered' configurations \cite{Braun:2015aa}$^{\bullet\bullet}$. Addressing this fundamental question of how cells manage to find and maintain `solutions' is challenging, at least in part, due to methodological limits in distinguishing relevant from irrelevant determinants distributed over broad spatial and temporal scales \cite{Gutenkunst:2007aa}$^{\bullet}$\cite{OLeary:2015aa}$^{\bullet}$. The presumed regulatory processes might operate at the levels of transcription, translation, protein folding and positioning, collective protein-protein interactions, protein degradation, protein kinetics and the biochemical modulations of these paths.  What makes membrane excitability particularly attractive to study in this context is the relatively sound understanding of the functional end-product, the action potential, and its amenability to experimental manipulations at both microscopic (channel protein) and macroscopic (membrane potential) levels. 

Emergence and maintenance of excitability is often considered in terms of arriving at and remaining about a manifold of `solutions' embedded in a high dimensional space composed of two families of parameters, structural and kinetic. In the Hodgkin and Huxley original model, the main structural parameters are membrane capacitance ($C_{M}$), and maximal sodium ($\bar{g}_{\text{Na}}$), potassium ($\bar{g}_{\text{K}}$) and leak ($\bar{g}_{\ell}$) conductances, and the many kinetic parameters are expressed as six transition rate functions: $\alpha_{n}(v)$, $\beta_{n}(v)$, $\alpha_{m}(v)$, $\beta_{m}(v)$, $\alpha_{h}(v)$, $\beta_{h}(v)$. The tables and figures exposed by Hodgkin and Huxley in their report \cite{Hodgkin:1952jt} suggest that kinetic parameters vary over a range of ca. $\pm 20\%$ and structural parameters over a factor of two or more. A glimpse to the nature of the manifold that supports excitability in the Hodgkin-Huxley parameter space is provided in Figure 1, where either pairs of parameters (Figure 1a) or the entire set of ten parameters (Figure 1b-c) are randomly perturbed over a moderate range ($\pm$25\% relative to the original HH values). Clearly, to maintain excitability a `coordinated tuning' of physically independent entities (different protein populations and their kinetic parameters) is required. For each pair of parameters there exists an `allowed' direction of change that is less prone to impact on excitability, and a `not-allowed' direction that is more prone to cause a qualitative modification of the macroscopic order. While imaginable in two-dimensional settings (Figure 1a), coordinated tuning becomes intricate when all parameters are considered simultaneously (Figure 1b-c). The parameter range apparent in the Hodgkin and Huxley's 1952 \textit{Journal of Physiology} paper is indicative to the fact that excitable cells actually utilize the space in maintaining functional order. But the most convincing evidence to that effect comes from a series of studies in the relatively well defined settings of the lobster stomatogastric-ganglion, showing that whether one observes the system from the point of view of a network or that of the single neuron, degeneracy of functional parametric configurations enables the emergence of excitability, as well as its maintenance in response to a rich repertoire of neuromodulatory effects and unforeseen environmental challenges \cite{marder2011variability}\cite{OLeary:2014qf}\cite{marder2014neuromodulation}$^{\bullet\bullet}$. 

\begin{figure*}[t!] \includegraphics[width=\textwidth]{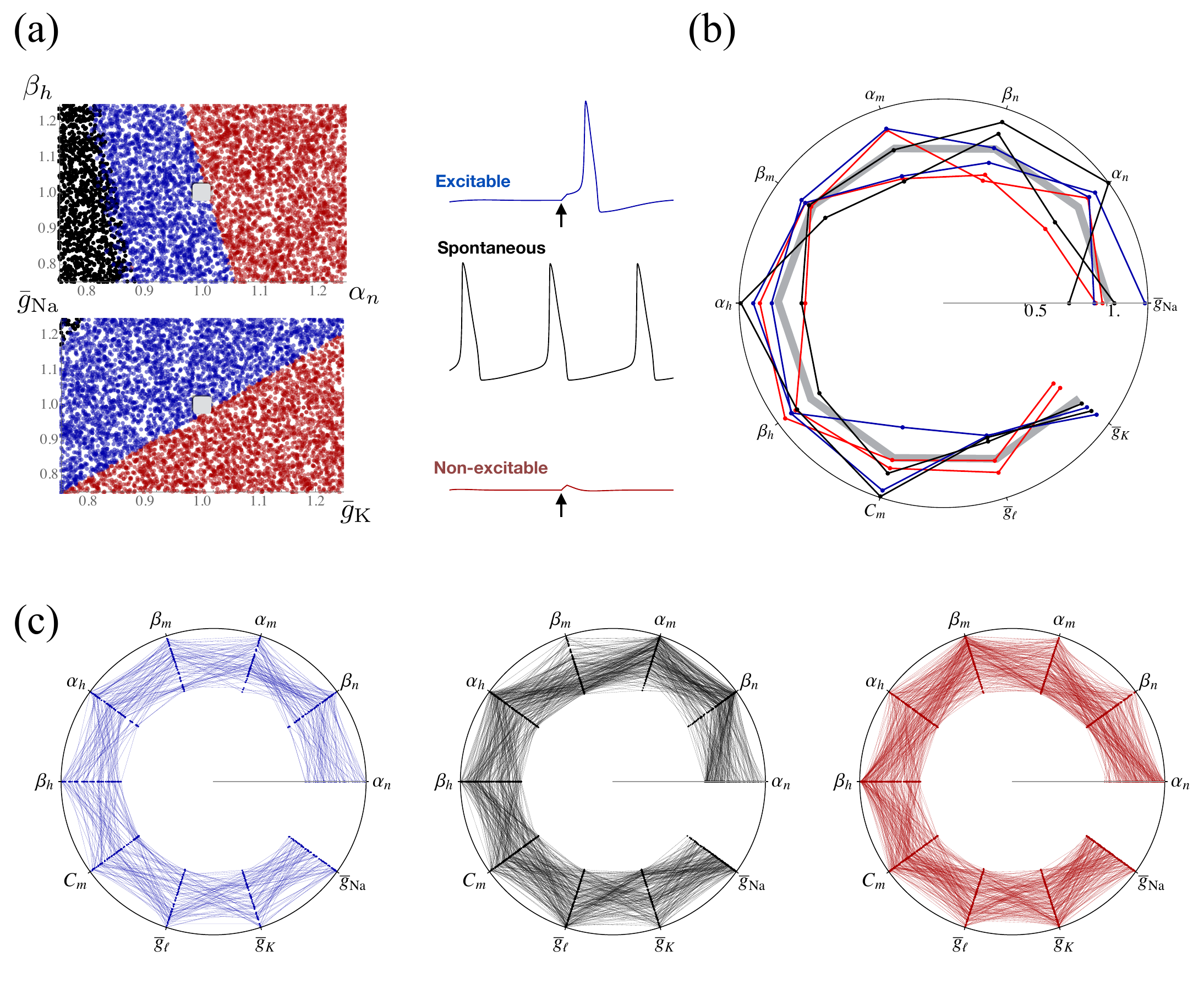} \caption{Manifold of excitability intuited. (a) Outcome of the original Hodgkin-Huxley model for membrane excitability with the values of the kinetic functions $\alpha_{n}(v)$ and $\beta_{h}(v)$ or structural parameters $\bar{g}_{\text{Na}}$ and $\bar{g}_{\text{K}}$ being scaled by randomly chosen factors over a fairly moderate range of 0.75 to 1.25 relative to their standard values (which are, interestingly, edgy; indicated by small grey squares in the middle of both examples). Classes of resulting excitability status were defined based on membrane response to a current stimulus (one millisecond long at x1.25 Hodgkin-Huxley original threshold stimulus) and depicted by color as indicated in the voltage traces to the right. In panel (c) the whole set of ten parameters was randomly perturbed 500 times over a range of 0.75 to 1.25 (that is, $\pm$25\%) relative to their standard values. Each set of parameters is presented as a line connecting the values in a polar plot; several examples shown in panel (b), where the Hodgkin-Huxley original parameter set is depicted by a thick grey line. Each parameter set was classified as giving rise to an excitable solution [panel (c), left], spontaneously spiking [panel (c), middle] and non-excitable [panel (d), right]. Note that all three excitability modes distribute throughout the range tested.} \end{figure*}

The manifold metaphor enriches the range of concepts available for description of mechanisms underlying the emergence and maintenance of excitability. Traditionally, we find ourselves adopting a \textit{control theoretical} framework to account for the coordination between values of different parameters, compensating `unwarranted' changes in one (or more) by intelligibly changing others \cite{OLeary:2011aa}.  The language used includes terms such as error signals, set-points, tuning-rules; these are straightforwardly mappable to the general concept of homeostasis, a bedrock of physiological thinking. Most studies related to the emergence and maintenance of excitability use the control theoretical framework to explain the coordination of structural parameters. These studies, extensively discussed in the literature (for instance \cite{OLeary:2014qf,marder2014neuromodulation}), stem from a general idea presented in the early 1990's, coupling channel protein expression to past electrical activity. Intracellular Ca$^{2+}$ concentration is the default choice for a leaky integrator of electrical activity because its influx through voltage-gated channels is enhanced by depolarizing membrane potentials; but the general message is independent of the actual identity of the integrator. Several modeling studies have since shown that such activity-dependent regulation mechanisms may robustly drive cells to become excitable. Moreover, activity-dependent regulation of protein expression levels was offered as an explanation for the different realizations underlying similar excitability modes, the `many-to-one' relations exemplified in Figure 1. In a recent elegant analysis the framework was further extended to account for experimentally observed patterns of correlation in the expression of different ionic channels, by (reasonably) assuming differential regulation of transcription \textit{rates} \cite{o2013correlations}.  Taken together with advances in understanding tricks used by cells to optimally control protein synthesis at large (e.g. \cite{Li:2014aa}), these and related studies mark a steady progress, providing experimental and theoretical grounds to believe that control-based tuning rules may intelligibly coordinate maximal conductances -- key \textit{structural} parameters underlying excitability.  

Yet, regulation of excitability by controlling densities of membrane channel proteins comes at a price. Even if we turn a blind eye to the entailment of order $N$~constraints for controlling an $N$-dimensional system, there is an issue that concerns timescales. Limited by rates of protein synthesis and degradation, it is not straightforward to picture activity-dependent control of channel protein densities as a mechanism covering regulation of excitability amidst dynamic constraints over the entire spectrum extending from sub-second to hours. In this context a recently budding stream of ideas suggests a potential contribution of \textit{kinetic} parameters to the emergence and maintenance of excitability over a broad range of timescales. The standard HH model assumes that the state transitions governing elementary reactions are all voltage dependent, operate at more-or-less the millisecond range and independent of each other. But as wisely acknowledged by Hodgkin and Huxley in a comment about the range and applicability of their equations (page 541 in \cite{Hodgkin:1952jt}), their assumptions do not hold when one is interested in dynamics beyond the timescale of a single spike. Even in the case of the voltage gated sodium conductance -- the very engine of excitability -- many years of extensive study (e.g.,~\cite{Toib:1998rr,Ellerkmann:2001bs,WernerUlbricht10012005,Marom:2010fk,Silva01032013a,Silva:2014aa}) show that channels fluctuate at much slower timescales (in an activity-dependent manner) between a pool of microscopic states that is available for the generation of excitability (depicted $A$), and a pool of microscopic states that is not available for such a process, depicted $(1-A)$. The set of available states is rather compact and may indeed be imagined as encompassing all the standard Hodgkin and Huxley transitions that are functionally proximal to the open state (Figure 2a). It is compact in the sense that the involved states are strongly coupled by rapid voltage dependent transitions with a characteristic time scale at the range of milliseconds. In contrast, the set of unavailable states $(1-A)$ is extended. It may reflect many distorted versions of the functional protein under conditions where the organization, otherwise enforced by hyperpolarized membrane potential, is compromised upon extensive depolarizing activity \cite{catterall2015deciphering}$^{\bullet\bullet}$. In the $(1-A)$ set, states are coupled by a mixture of weak voltage-dependent and voltage-independent transition rates with time scales ranging from milliseconds to many minutes. In accordance, residence times in the unavailable pool are broadly distributed -- maybe scale invariant \cite{Toib:1998rr,Ellerkmann:2001bs,Goychuk:2004kx,millhauser1988diffusion,millhauser1988rate} -- reflecting the depth of the population distribution in that pool (Figure 2b). This implies a potential to become dormant in an activity-dependent manner for a duration ranging from tens of milliseconds to many minutes and possibly hours. Thus, regulation of excitability needs not be solely based on changes to the expressed structure -- i.e., the actual number of proteins residing in the membrane. Rather, an effective balance between channel populations amidst dynamic constraints may also be achieved by a process acting on already expressed proteins. This interpretation is potentially attractive to explore. It partially relaxes the challenge of tightly controlled structural parameters, enables local regulation of excitability in complicated oxo-dendritic morphologies, and opens a wide range of potential excitability configurations for a cell to resume while interacting with rich and dynamic input in multiple time scales. 

\begin{figure*}[t!] \begin{center} \includegraphics[width=\textwidth]{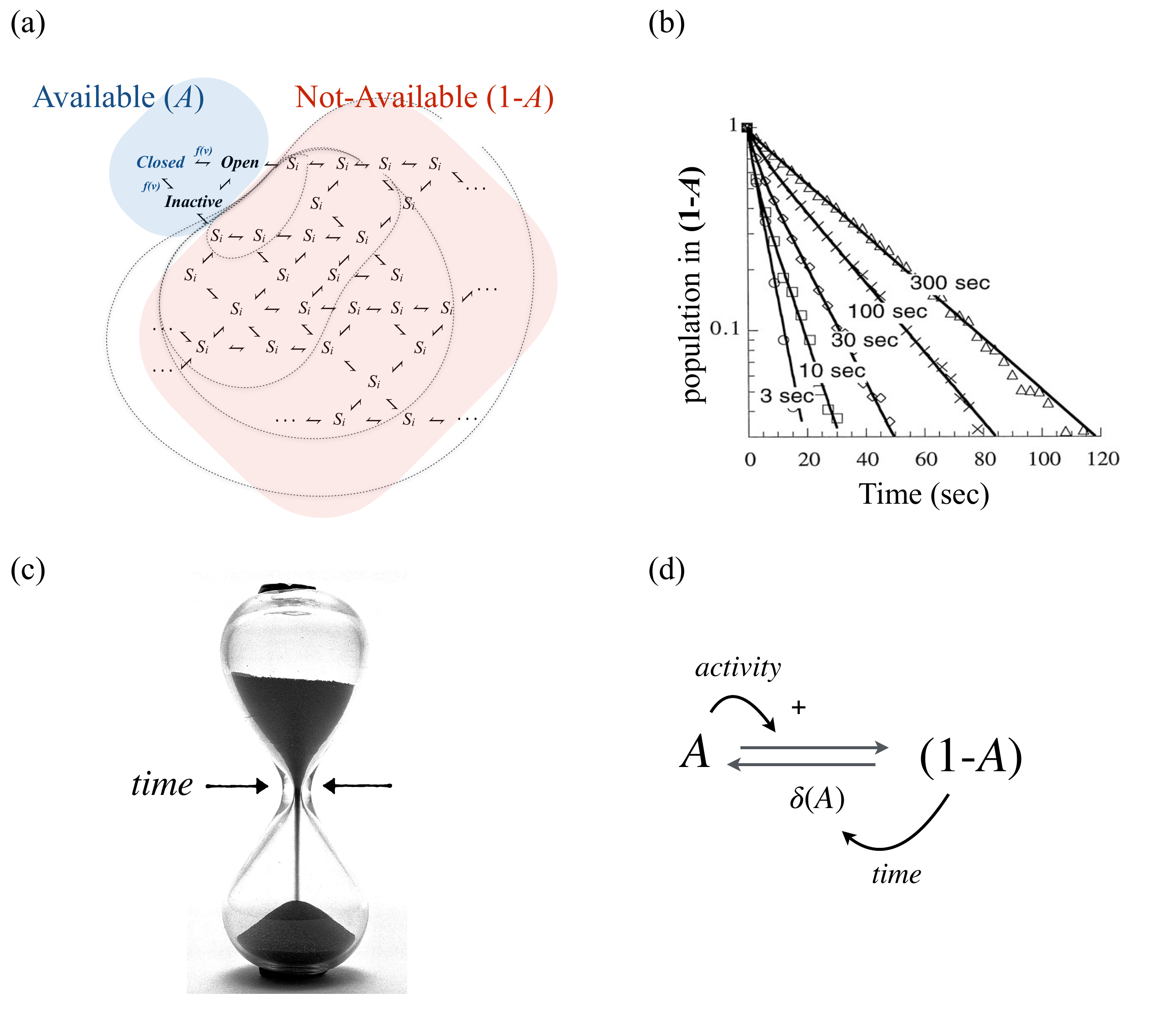} \caption{Adaptive rate model. (a) The state-space of voltage-gated ionic channels is divided to available and unavailable state populations; the longer the history of activity the deeper the channel drifts into the unavailable pool; the time constant for recovery into the available pool scales accordingly. This is shown in panel (b), where recovery from inactivation in sodium channels is plotted as a function of duration of past activity (modified from \cite{Toib:1998rr}). (c) The behavior shown in panel (b) is analogous to the neck of an hourglass being modified by the time passed since it was put on its head. (Image L0011338, CC BY 4.0 [http://creativecommons.org/licenses/by/4.0], via Wikimedia Commons.) (d) Schematization of adaptive rate kinetics.} \end{center} \end{figure*}

There are many experimental indications for channel-level broadly distributed timescales in slow adaptive dynamics of excitability (e.g.,~\cite{fairhall2001efficiency,pozzorini2013temporal}), but here we are interested in ways by which these complex kinetics contribute to regulation of excitability. One suggestive experimental support to this effect comes from long-term recordings of response dynamics in detached cortical neurons \textit{in-vitro}, using extracellular stimulation (to prevent damaging the cell) and applying careful measures to ascertain experimental stability \cite{AsafGal12012010,gal2013entrainment,reinartz2014synaptic}. Such long time series of evoked spikes (minutes to hours) are dominated by scale-invariant rate statistics. When closely examined, a given neuron seems to wander around, \textit{reversibly} visiting different activity signatures, changing its excitability status as if randomly switching identities between reliable, irregular, bursting or compound response patterns. The reversible nature of this behavior, as well as the durations of residence in these different excitability modes, preclude changes in membrane protein densities. Such quasi-stable long-term modes (under constant membrane channel densities) may be accounted for by assuming interaction between slow channel kinetic parameters and stochasticity; a model of such flavor nicely fits experimental traces \cite{Soudry:2014aa}. However, the huge space of unavailable states within which channels diffuse and drift due to electrical activity may offer more: it provides a bed for the emergence and maintenance of self-organized, adaptive excitability modes. An example for effective mapping between rich ionic channel kinetics, adaptivity and self-organization, involves a low dimensional mathematical expression of channel population gating -- the \textit{adaptive rate} model \cite{marom2009adaptive}. The general idea is that activity impacts on a global recovery rate from the unavailable pool of states as if an hourglass neck narrows down as a function of time past since the glass was put on its head (Figure 2c--d). Such a process may be expressed by a logistic-function-like form $\dot{A} = -f(\gamma) A + g(A)(1-A)$ that captures the essence of the $A\leftrightarrow(1-A)$ dynamics, where $f$ is a function of the neural activity measure $\gamma$, and $g(A)$ is a monotonically increasing function of the population size occupying state $A$. As recently pointed out \cite{gal2013self}$^{\bullet\bullet}$\cite{excitability2014single}, this model is formally identical to a mean-field expression of a second order phase transition in a global contact process \cite{dickman2000paths}.\footnote{A model of probabilistically interacting elements much used in epidemics.} Taken together with a natural feedback that couples between available population ($A$) and excitability, the adaptive rate model gives rise to self-organized positioning of the system about an edge that separates between excitable and non-excitable phases, and nicely corresponds to experimental observations \cite{gal2013self,excitability2014single}. Experimental analyses show that as one approaches that edge of excitability \cite{gal2013self,excitability2014single}\cite{Meisel:2015aa}$^{\bullet\bullet}$, response fluctuations become larger and slower, tempting to propose a link between the large space of channel states and the power-law statistics of firing rate. Often, being edgy is an edge; it enables small changes to significantly impact on system behavior. In this context it is admittedly difficult to avoid pointing the reader's attention to several similarities between the above cluster of theoretical and experimental arguments, and the toy model of self-organized criticality \cite{dickman2000paths,dickman1998self,hesse2015self,schuster2014criticality}, reservations acknowledged \cite{Bedard:2006aa,Touboul:2010aa,beggs2012being}.

To summarize, advancements in the study of emergence and maintenance of excitability were described. Alongside ongoing studies that focus on control-based regulation of channel protein expression, a recent interest in kinetic-based self-organization of excitability was pointed to, where the large space of protein states and activity-dependent transitions between these states is perceived as a bed for the emergence and adaptation of excitability modes. One might picture the emergence and maintenance of excitability as driven by a compound machinery: structural changes that position the cell in a `ballpark' within which there are sufficient degrees of freedom to move between activity signatures over a wide range of timescales, reflecting activity-dependent changes in kinetic parameters. The former operates on the timescale of hours or more, whereas the latter caters to changes at shorter physiological scales. Given this `continuum' of timescales we might wish to extend our standard models of excitability and treat these structural and kinetic values as dynamic variables rather than parameters, at least when dynamics beyond the time scale of a single action potential are considered.

The study of excitability, from Galvani and Helmholtz, through Adrian, Hodgkin and Huxley, Neher and Sakmann, all the way to McKinnon's elucidation of a channel protein structure, should no-doubt make us physiologists very proud of our discipline. But there is still a way to go. Present concepts and technologies make it possible to implement the biophysical understanding of excitability in the more general context of mechanisms underlying the establishment and maintenance of functional cellular organization. In doing so, whether we adopt the path of control theory or the statistical physics one, we should bear in mind that the full story of the emergence and maintenance of excitability must also include a relational aspect. As commented elsewhere, pertaining to the wider context of behavioral and brain sciences \cite{marom2015science}, searching for the coordinates of a complex process by focusing solely on what goes on inside the object of analysis is analogous to searching for a shape of a mug in raw clay; the latter may be used to create one, but there is no mug in the raw material; it takes interaction with a `potter environment' to make one. The interactions between the cell and the tissue (network) within which it is embedded are critical determinants in moulding an instantiation of a given cellular configuration out of the large space of all possible configurations. Experimentally analyze and theoretically formulate the nature of relational processes is no less than the challenge of physiology at large, a challenge we are still falling short in meeting due to present conceptual and methodological (to be distinguished from technological) limitations. 

\newpage
\bibliographystyle{unsrt} 
\bibliography{SMbib}

\subsubsection*{Annotations for references}

\begin{itemize}

\item[$^{\bullet\bullet}$] E. Braun. The unforeseen challenge: from genotype-to-phenotype in cell populations. Reports on Progress in Physics, 78(3):036602, 2015:  A view of cell-state organization as a dynamical process, where the genome provides a set of constraints on the spectrum of regulatory modes, analogous to boundary conditions in physical dynamical systems.
\item[$^{\bullet}$] M.K. Transtrum, B. Machta, K. Brown, B.C. Daniels, C.R. Myers, and J.P. Sethna. Perspective: Sloppiness and emergent theories in physics, biology, and beyond. J. Chem. Phys., 143, 2015: A recent perspective on the concept of ``sloppiness'', where many parameter sets can exhibit the same behavior, in natural systems in general, and in biological systems in particular. Impacts on effectiveness of statistical and computational tools are described. A most inviting discussion section.
\item[$^{\bullet}$] T. O'Leary, A.C. Sutton, and E. Marder. Computational models in the age of large datasets. Curr. Opin. Neurobiol., 32:87--94, 2015: Points to challenges entailed by the ``curse of richness'' -- the unfathomable sea of data made available by technological advances in experimental neuroscience. Suggests being open to more loosely constrained conceptual models that explore broad hypotheses and principles.
\item[$^{\bullet\bullet}$] E. Marder, T. O'Leary, and S. Shruti. Neuromodulation of circuits with variable parameters: Small circuits reveal principles of state-dependent and robust neuromodulation. Ann. Rev. Neurosci., 37:329--346, 2014: An extensive source of references bringing together a host of data and computational results pertaining to variability, degeneracy and adaptation at the cellular and network levels, integrated with responses to neuromodulation. 
\item[$^{\bullet\bullet}$] W. A. Catterall and N. Zheng. Deciphering voltage-gated Na+ and Ca2+ channels by studying prokaryotic ancestors. Trends in Biochemical Sciences, 40(9):526--534, 2015: Accessible integration of recent understanding of sodium and calcium ionic channels, based on high-resolution structural insights from prokaryotic channel proteins.
\item[$^{\bullet\bullet}$] A. Gal and S. Marom. Self-organized criticality in single-neuron excitability. Phys. Rev. E, 88(6):062717, 2013: Experimental and theoretical arguments, at the single-neuron level, mapping neuronal response fluctuations to a process that positions the neuron near a transition point that separates excitable and unexcitable phases. 
\item[$^{\bullet\bullet}$] C. Meisel, A. Klaus, C. Kuehn, and D. Plenz. Critical slowing down governs the transition to neuron spiking. PLoS Comp. Biol., 11(2):e1004097, 2015: Studying the transition from neuronal quiescence to spiking, these authors demonstrate the membrane's tendency to recover more slowly from perturbations upon approaching its transition point. Interpreted in terms of critical phenomena.

\end{itemize}

\end{document}